**Title**

Non-invasive detection of nanoscale structural changes in cornea associated with cross-linking treatment

Yi Zhou[1], Sergey Alexandrov[1], Andrew Nolan[1], Nandan Das[1], Rajib Dey[1], and Martin Leahy[1,*]

[1] National University of Ireland, Tissue Optics and Microcirculation Imaging Facility, National Biophotonics and Imaging Platform, Galway, Ireland

*Corresponding Author:
 E-mail: martin.leahy@nuigalway.ie

**Abstract**

Corneal cross-linking (CXL) using UVA irradiation with a riboflavin photosensitizer has grown from an interesting concept to a practical clinical treatment for corneal ectatic diseases globally, such as keratoconus. To characterize the corneal structural changes, existing methods such as X-ray microscopy, transmission electron microscopy (TEM), histology and optical coherence tomography have been used. However, these methods have various drawbacks such as invasive detection, the impossibility for *in vivo* measurement, or limited resolution and sensitivity to structural alterations. Here, we report the application of over-sampling nano-sensitive optical coherence tomography (nsOCT) method for probing the corneal structural alterations. The results indicate that the spatial period increases slightly after 30 minutes riboflavin instillation but decreases significantly after 30 min UVA irradiation following the Dresden protocol. The proposed non-invasive method can be implemented using existing OCT system, without any additional components, for detecting nanoscale changes with the potential to assist diagnostic assessment during CXL treatment, and possibly to be a real-time monitoring tool in clinics.

**1. Introduction**

Corneal ectasia can severely impair vision, especially in the progressive form caused by the inherent structural weakness of the cornea. [1, 2] Keratoconus, the most common form of corneal ectasia affecting nearly 1 in 375 individuals globally, is an ocular disorder characterized by corneal degeneration due to corneal thinning that results in a focally decreased radius of corneal curvature and abnormal wavefront aberrations. [3-5] Keratoconus is a non-inflammatory, progressive disease resulting in corneal instability in the structure. Sometimes it ultimately may require a corneal transplant. In recent years, a non-surgical keratoconus treatment, corneal cross-linking (CXL), was proved to be an effective way in halting the progression of keratoconus, meaning patients can avoid a corneal transplant. [6-9] Corneal cross-linking, using a combination of ultraviolet-A (UVA) light with 365 nm central wavelength and a photosensitizer (riboflavin, vitamin B2) for the procedure, shows a physiologic principle of tissue biomechanical modification. [10-12] CXL, which presents to be a minimally invasive treatment to stiffen the corneal stroma by strengthening the connections between collagen fibrils, was first approved by the U.S. FOOD and Drug Administration (FDA) in April 2016. [13]

Despite being in clinical use for several years, some of the principles and underlying processes, such as the role of oxygen, the optimal treatment time and the way to monitor and assess the treatment, are still being worked out. [14, 15] The basic theory behind this approach is photo-polymerization, which leads to the creation of chemical bonds between collagens and proteoglycans and other proteins within the corneal stroma. Various researches have been reported that the increase of tissue stiffness is dependent on the modifications of mechanical properties, which have a strong relationship with the ultrastructural alterations within the cross-linked cornea, such as the changes of the characteristics of collagen fibrils. [16, 17] Detection of structural changes in corneal stroma during the cross-linking procedure would allow a more detailed investigation of healing processes or the efficacy of drugs. Characterizing the structural changes could allow in the future to improve optical monitoring and improve treatment methods to repair damages in the cornea. [18-22] Transmission electron microscopy (TEM) and scanning electron microscopy (SEM) were used to study the relationship between mechanical behaviour and changes in the corneal ultrastructure after cross-linking in rabbit eyes. [23, 24] Although the nanoscale structure can be clearly imaged, these technologies will only suit the samples in vitro. Another option for detecting ultrastructure in the cornea is histology, which requires the anatomy of the tissue and therefore only applicable for *ex vivo* tissue. Some other methodologies including X-ray microscopy [25] and confocal microscopy [26, 27] also have various limitations, such as harmfulness from radiation damage, small imaging area or shallow imaging depth, causing difficulties in clinical eye diagnostics .

Optical coherence tomography (OCT), first introduced by its name in 1991 for the visualization of alterations in the human eye by a team in Massachusetts Institute of Technology (MIT), originates from optical low-coherence interferometry and generates 2D or 3D cross-sectional tomographic images by detecting the echo time delay and intensity of light which is reflected or back-scattered from internal structures in tissue. [28-30] OCT has proved to be of significant value in the visualization of transparent, weakly scattering media, such as the cornea, and obtained a substantial impact in the ophthalmology community. [15, 31-33] Efforts made in the last years in OCT technology have also made it possible to image highly scattering tissues, such as in dermatology, gastroenterology, oncology and interventional cardiology. Given its high-resolution, noninvasive, fast-speed and three-dimensional method, OCT fills the gap between the optical microscopy and currently available clinical imaging methods considering both imaging resolution and depth.

The interest in OCT for ophthalmic applications has rapidly increased in the past years. Refractive surgery on the cornea benefits very much from the existing OCT technology and is also used for phakic intraocular lens implantation. Laser-assisted in situ keratomileuses (LASIK) enhancement and lamellar keratoplasty are other applications using OCT technology. In all cases, OCT opened promising therapeutic and diagnostic options in both research and clinical applications in ophthalmology. The spatial and axial resolution of OCT is, however, also limited by the wavelength of the used light source and the used optical components. However, for specific treatments of the eye's cornea, the monitoring of cross-linking process requires image resolution in the nanometer range as documented using TEM and SEM. [23, 34]

In this study, we have presented the application of over-sampling nano-sensitive optical coherence tomography (nsOCT), which is proposed to retain the high spatial frequency information in the interference spectra, to probe the structural alterations inside *ex vivo* bovine cornea during CXL treatment with nanoscale sensitivity. The capability of the proposed nsOCT method to detect nanoscale difference has been proved experimentally by distinguishing two periodic Bragg grating samples with known dimensional sizes. The investigation of corneal cross-linking was performed using the Dresden protocol, which is the gold standard for the CXL treatment. The data of corneas were recorded and analyzed in three time nodes: 1) the epithelium removal, 2) a 30 min riboflavin instillation, and 3) a 30 min UVA irradiation. We present the over-sampling nsOCT B-scans and *en face* results at different corneal depths, showing a consistent consequence. The spatial periods of cornea increased slightly after 30 min riboflavin instillation but decreased significantly after 30 min UVA irradiation. The results suggest that the over-sampling nsOCT can be used to detect nano-sized structural changes valuable for corneal treatment methods.

## 2. Methods and Materials

### 2.1. Principle and demonstration of the over-sampling nsOCT

Scattering potential of a 3D object can be reconstructed from the distribution of the complex amplitude of the light scattered by the objective in far zone via 3D inverse Fourier transform. [35] Based on the theory of spectral encoding of spatial frequency (SESF) approach, [36-39] which was demonstrated for quantitative characterization of the structure with nanoscale sensitivity, nsOCT has been proposed to be a label-free depth-resolved sensing technique to probe structural changes at the nanoscale. [40-42] Later nsOCT has been used for study of the nanometer-scale structural changes of the human tympanic membrane in otitis media. [43] Recently a version of the nsOCT based on correlation between axial spatial frequency profiles has been published. [44] In this study, oversampling is used to increase the number of points in the reconstructed axial spatial frequency/period profiles. In conventional OCT, high-frequency information has been killed during the inverse Fourier transform, and the axial resolution is limited in micron level depending on the spectral bandwidth of the light source.

The nsOCT method carries the spectrally encoded axial spatial period from the $k$-space to the reconstructed object space by encoding each axial spatial frequency with one particular wavelength. The signal processing for nsOCT is illustrated in a flowchart in Figure 1. To realize over-sampling nsOCT, we first obtain the processed interference spectra after basically preliminary modifications, including $k$-space linearization, background noise removal, apodization, and dispersion compensation. Then the processed spectra are decomposed into several sub-bands using Tukey windows as for nsOCT imaging. To minimize the spatial frequency range for one sub-band, we then oversample the spatial frequencies by shifting the window with the same width, i.e. the starting point for each processing shifts along spatial frequency, as illustrated in Figure 1 (a). The FFT of each sub-band signal has been performed to get the amplitude along all the imaging depths for one spatial frequency range, as shown in Figure 1(b). By combining all the over-sampling spatial profiles, it is possible to reconstruct the dominant structural size of each voxel by counting the medium spatial periods with the maximum energy contribution, as indicated in Figure 1(c).

The collected complex amplitudes of the spectrum are converted to complex amplitudes of spatial frequencies ($V_z$) based on the following relation:

$$V_z = \frac{2n}{\lambda} \quad (1)$$

where $n$ is the refractive index, and $\lambda$ is the wavelength. The spatial periods ($H_z$) of the structure can be expressed as follows:

$$H_z = \frac{1}{V_z} = \frac{\lambda}{2n} \quad (2)$$

The nsOCT image can be then constructed by mapping the corresponding dominant spatial period to each voxel with an optimized threshold.

For further experimental validation of the proposed nsOCT method, we used two reflection Bragg gratings with periodicities of 431.6 nm and 441.7 nm (Corp. OptiGrate, Florida, United States) as the samples. Each periodic layer was fabricated with sinusoidal refractive index varies 1.483 ± 0.001. No variation in periodicity vs time if temperature of usage is <400°C (the period will be stable better than 1 pm). With the purpose of mimicking an imaging condition similar to real tissue, a two-layer plastic tape were placed on the top of the Bragg gratings and a tissue paper, as shown in Figure 2(a).

Figure 2(a) and 2(b) show the conventional intensity-based OCT B-scan and *en face* images, where the structural difference between the Bragg gratings cannot be recognized since the high-frequency information is not available using general OCT processing. While following the proposed nsOCT method, both the nsOCT B-scan and *en face* images were constructed with spatial periods mapping as presented in Figure 2(c) and (d), easily visualizing the nanometer structural difference (~10 nm) between two Bragg gratings. The spatial periods of the tape and tissue paper are quite diffused with various values as the structural sizes of them are randomly distributed. From the proof-of-concept demonstrations, we successfully verify that the nsOCT method is a promising way to sense nanoscale structural alterations along the depth axis.

Further, to clarify the difference between over-sampling nsOCT and the previous nsOCT method, we have added both simulation and experiment results. In the simulation, we produce seven samples with different axial periodicities of 425 nm, 428 nm, 431 nm, 434 nm, 437 nm, 440 nm and 443 nm. As shown in Figure 3, the structural difference of 3 nm can be distinguished by the over-sampling nsOCT. However, nsOCT without over-sampling can only detect the difference of 6 nm but cannot distinguish 3 nm difference. In the experiment, we still use the Bragg gratings for the demonstration. For the Bragg grating with 431.6 nm periodicity, we measure 430.4 nm (error: 1.2 nm) by over-sampling nsOCT while 428.2 nm (error: 3.4 nm) by the previous nsOCT method, as shown in Figure 4. It also demonstrates that nsOCT after over-sampling can provide a higher measurement accuracy. Another result of the Bragg grating with 441.7 nm periodicity is attached in the supplementary files Figure S1, also showing an improved sensitivity after over-sampling process.

### 2.2. Ex vivo bovine corneal cross-linking experiment

In the CXL experiments, we chose bovine corneas for more consistent sample quality from the local abattoir. The bovine eyes were harvested in the evening of slaughter from a local abattoir (Athenry Quality Meats Ltd, County Galway, Ireland) within 2 hours post mortem and stored at 4 °C. The abattoir granted permission for the eyes to be used in research. Eye globes with intact epithelium and clear cornea were selected for the *ex vivo* studies. As it is an epi-off cross-linking treatment, we first removed the epithelium with a stainless-steel spatula. The CXL process was performed in a dark environment using riboflavin (MedioCROSS M, 0.1% Riboflavin, 1.1% HPMC) as the photosensitizer and final irradiation with UV-A light (UV-XTM 2000, IROC Innocross, Zurich, Switzerland) with a illumination spot size of 9.5 mm, 3.0 mW/cm$^2$ intensity, and energy dose of 5.4 J/cm$^2$.

Following the Dresden protocol, which is the current gold standard of CXL treatment, we conducted the structural detections at the three treating steps including immediately after epithelial removal, after 30 min riboflavin instillation, and after 30 min UVA irradiation with a further continuous installation every 2 minutes.

For the data acquisition, a commercial spectral-domain optical coherence tomography (SDOCT) system (Telesto III, Thorlabs, Inc. New Jersey, United States) with an objective lens LSM03 (NA=0.055, lateral resolution=13 μm) was used in our experiments. This high-resolution OCT system, operating at the central wavelength of 1300 nm with sensitivity 96 dB @76kHz rate, can reach an axial resolution of 5.5 μm and imaging depth of 3.6 mm in air.

## 3. Results

We used a laser scanning confocal microscopy system (FV 3000, Olympus ) to investigate the differences between the untreated and treated corneas. We used a 405 nm excitation wavelength and an objective lens with an NA of 0.85 (UPLSAPO 20XO) for the imaging process. The confocal microscopy images of the virgin and treated corneas are presented in Figure 5. The virgin corneas consisted of highly reflective and well-demarcated cell nuclei with an oval shape, indicated by white arrows. Neither keratocyte processes nor collagen fibers could be visualized. However, the crosslinked cornea was populated with some reflective interconnected stellate structures. These structures contained elongated nuclei and keratocyte apoptotic bodies. The confocal microscope images show that the corneal crosslinking has occurred during the CXL treatment.

Figure 6(a, c, e) shows the intensity-based OCT B-scans for the three processing steps, respectively. However, the conventional OCT images cannot identify any changes smaller than the resolution limit and no structural differences inside corneas can be identified. Here, we applied the proposed nsOCT method for analyzing the structural responses during CXL treatment, showing B-scans with spatial periods mapping, as presented in Figure 6(b, d, f). The nsOCT B-scans show that there are no observable structural alterations between the virgin cornea and the cornea after 30 min riboflavin instillation, only presenting a minor increase after riboflavin immersion. But the spatial periods had an obvious decrease after 30 min UVA irradiation, meaning that the structural size decreased significantly after the CXL treatment. In Figure 6(f), the spatial periods after CXL process were clearly detected to be smaller by several nanometers compared to Figure 6(b, d).

To quantitatively demonstrate the structural variations associated with CXL treatment, we also investigated the histograms and boxplots of spatial period distributions for B-scans at all the three steps. Figure 7(a, b, c) indicates the spatial period results for one of the bovine corneas. The mean spatial periods of the virgin cornea, the cornea after 30 min riboflavin instillation and after 30 min UVA irradiation were measured to be 657.1 nm, 658.1 nm and 649.6 nm, respectively, illustrating the slightly larger spatial size after riboflavin immersion but substantially smaller after the CXL procedure. In the meantime, the histograms of spatial period distributions also show that the treated cornea has shifted to be smaller structural sizes.

Further, we have conducted a series of repeating experiments on different bovine eyes under the same CXL protocol and signal processing. Analyzing the experimental data of ten bovine eyes associated with CXL treatment, we made a summary on both the intensity and spatial period shifts as presented in Table 1 and Table 2, respectively. We found that the intensity values fluctuated both negatively and positively with no continuous change in one single way, i.e., it is not possible to detect the variations during the CXL process by conventional intensity-based OCT. Additionally, we included the intensity histograms of the virgin cornea, the cornea after 30 min of riboflavin instillation, and the treated cornea in Figure S2. Again, we cannot observe any obvious intensity shifts between them.

Summarizing all the experimental sets using the proposed over-sampling nsOCT method, we evaluated that the spatial period has changed from 653.9± 1.7 nm (mean ± standard deviation) of the virgin corneas to 647.5± 3.3 nm of the treated corneas. The spatial periods inside the cornea have decreased -6.4± 2.8 nm with the percentage of (-0.98± 0.43)% after CXL treatment. Furthermore, one can also notice that there is a consistent decrease of spatial period for all the bovine corneas in the experiments, providing a reasonably convincing verification of the nanoscale structural changes associated with the CXL treatment.

The acquisition of 3D volumetric data was used to obtain *en face* images along with different imaging depths. Each *en face* image was obtained over an area with 2 mm x 2 mm size and 400 x 400 A-lines. The treatment was performed with the same protocol as described above. The conventional OCT *en face* images (Figure 8a-c) is not able to resolve any features that could indicate any changes due to the CXL procedure. Nonetheless, the nsOCT *en face* images could visualize the nanoscale alterations especially between the virgin cornea and cross-linked cornea, showing a minor spatial period after the treatment, as indicated in Figure 8(d, e, f). In the result, the mean spatial periods of *en face* data of the virgin cornea, the cornea after 30 min riboflavin instillation and the cross-linked cornea were calculated to be 653.2 nm, 654.3 nm and 648.6 nm, respectively. To evaluate the change of the nsOCT *en face* images at different depths, a series of scans at increasing depth were assembled in Video S1.

Figure 9 shows the spatial period (Y-Axis) along a depth (X-Axis) from 200 to 770 µm computed at depth intervals of 38 µm step width. The change of the spatial period for each treatment step is easily recognizable in Figure 9. The spatial periods for a virgin bovine cornea (red circles) are consistently smaller than those for a cornea treated with riboflavin (yellow squares). A significant change occurs after the 30 minutes of UVA irradiation and the values of the spatial periods reduce by about 3 nm from 652 nm to 649 nm.

Based on the values plotted in Figure 9, a paired t-test was performed (samples size ten per group) which confirmed the high significance ($p<10^{-4}$) between the virgin corneas and the group after 30 min riboflavin instillation. The p-value test for spatial changes between the virgin and CXL treated corneas provided an even larger significance $p<10^{-13}$.

## 4. Discussion

The experimental results of Bragg gratings with known structural sizes have demonstrated the feasibility of the over-sampling nsOCT method to distinguish the nanoscale structural alterations in both B-scan and *en face* images. In the study of *ex vivo* bovine corneas associated with CXL treatment, both the conventional intensity-based OCT images and the corresponding nsOCT images were constructed for the virgin cornea after epithelial removal, the cornea after 30 min riboflavin instillation and after 30 min UVA irradiation. From the conventional OCT images, we can hardly recognize the internal structural changes of the cornea due to the limited axial resolution of several micrometers. Nonetheless, our investigation using the nsOCT modality could detect the ultrastructural variations inside the cornea during CXL treatment, showing that the spatial periods slightly increased after 30 min riboflavin instillation but significantly decreased after 30 min UVA irradiation.

According to the existing reports, the structural size inside the cornea should decrease after riboflavin absorption due to the dehydration from the hypertonic photosensitizer solution. In our study, we found an increased spatial period after 30 min riboflavin instillation in a dark circumstance. We think the reason why we got opposite results with TEM or SEM method is because of the increase of refractive index inside cornea after riboflavin immersion. In OCT system, we detect the optical path difference that depends on both the physical structural size and the refractive index of sample. The minor increment of the spatial period after 30 min riboflavin instillation may be related to the higher refractive index of riboflavin compared to the cornea itself. The mean refractive index inside the stroma will hereby increase when riboflavin penetrates inside the cornea, resulting in a larger spatial period.

To investigate the interfibrillar distance and fibril diameter changes of corneal stroma after CXL treatment, a few methods, such as TEM, SEM, and X-ray imaging, have been discussed in the previous literature. T. Sibillano et al [25] studied on the *ex vivo* bovine corneas using small-angle X-ray scattering (SAXS) microscopy and revealed that the interfibrillar distance decreased from ~62 nm (virgin cornea) to ~56 nm (treated cornea). The results also concluded the decrease in the fibril shell thickness from ~11 nm to ~9 nm after CXL treatment because of the effect of cross-linking within proteoglycan core proteins, resulting in a decreased spatial periods. The similar results on structural changes associated with CXL process were also reported by Ho, Leona TY et al, [45] Freund et al [46] and Cheng et al, [47], showing a good agreement with the proposed nsOCT method. More results have also been reported on the rabbit, porcine and sheep corneas. For instance, Sally Hayes [20, 24] found that the fiber diameter of in vitro porcine cornea increased while the interfibrillar spacing decreased. *Ex vivo*, CXL treatment on sheep corneal tissue samples was associated with slight decreases of collagen intermolecular spacing.

Further, dehydration may be one of the factors contributing to the structural changes during CXL treatment, especially for *ex vivo* samples. Therefore, to ensure that the structural alterations are induced by corneal cross-linking treatment other than some other factors, such as dehydration, a control experiment was arranged. The bovine eye was treated in the same CXL procedure as mentioned before, except that half of the cornea was blocked from the UVA illumination.

The virgin cornea was imaged immediately after removing the epithelium, as Figure 10(a) shows the intensity-based OCT image and Figure 10(b) presents the nsOCT image. Then the riboflavin was dropped on the entire cornea for 30 minutes in a dark circumstance. In the last step, different from the previous process, a blocking long pass filter was used to stop the UVA illumination in the right half of the cornea. As presented in Figure 10(c), the CXL treatment will only occur in the left half of the cornea, as the UVA light cannot reach the right half. We calculated that the spatial period of the left half was 647.9 nm while the right half is 653.6 nm, also showing a decreased spatial period after CXL treatment. The corresponding nsOCT image, shown in Figure 10(d), demonstrated the nanoscale structural differences between left and right half of the same cornea.

## 5. Conclusion

In this study, we reported that the nano-structural changes inside *ex vivo* bovine cornea associated with CXL treatment can be clearly detected by the proposed over-sampling nsOCT method. Multipe experiments on several groups of samples have consistently shown that the spatial periods inside corneal stroma increased slightly after 30 minutes riboflavin instillation but decreased significantly after 30 min UVA irradiation. *En face* nsOCT images at different corneal depths have also confirmed the consistent consequences, demonstrating the nanoscale structural decrease after the CXL treatment.

To the best of our knowledge, there is no clear tools or procedures to follow up with the patient corneal monitoring during CXL treatment. The currently preferred procedure is the epithelium mapping which reflects the behaviour of the other layers of the cornea, but it is not capable of the epi-off surgery or real-time ultrastructural inspection. The proposed method can be implemented using the existing OCT system, without any additional components, therefore it can be relatively straightforward to be applied in *in vivo* tissues and translated to clinical use as a novel imaging system. Due to its fast, non-invasive detecting method and nanoscale sensitivity, this unique technology is potential to be an indicator in diagnostic assessment associated with CXL treatment, and possibly to be a real-time monitoring tool in clinics as a fast way to receive feedback from patient's tissue. Future work will aim to implement this method for *in vivo* corneal detections associated with CXL treatment, including monitoring the nanoscale structural variations at different treating step and also the postoperative assessment.

**Acknowledgements**

We would like to thank Cerine Lal, Anand Arangath, Ryan McAuley and Kai for their help with the fruitful discussions. This project has received funding from the European Union's Horizon 2020 research and innovation programme under grant agreements no. 761214 and no. 779960. The materials presented and views expressed here are the responsibility of the author(s) only. The EU Commission takes no responsibility for any use made of the information set out. Nandan Das received funding from Irish Research Council (IRC), under Government of Ireland postdoctoral fellowship with project ID: GOIPD/2017/837. The authors also acknowledge the facilities and scientific and technical assistance of the Centre for Microscopy & Imaging at the National University of Ireland Galway.

**CONFLICT OF INTEREST**

The authors declare no financial or commercial conflict of interest

**SUPPORTING INFORMATION**

Additional Supporting Information may be found online in the supporting information tab for this article.

**Figure S1:** Experimental results of the Bragg grating with axial periodicity of 441.7 nm by previous nsOCT and the over-sampling nsOCT method.

**Figure S2:** Intensity histograms of the virgin cornea, the cornea after 30 min of riboflavin instillation, and the treated cornea.

**Video S1:** Over-sampling nsOCT *en face* images of the virgin cornea and treated cornea at different depths.

**FIGURES AND TABLES**

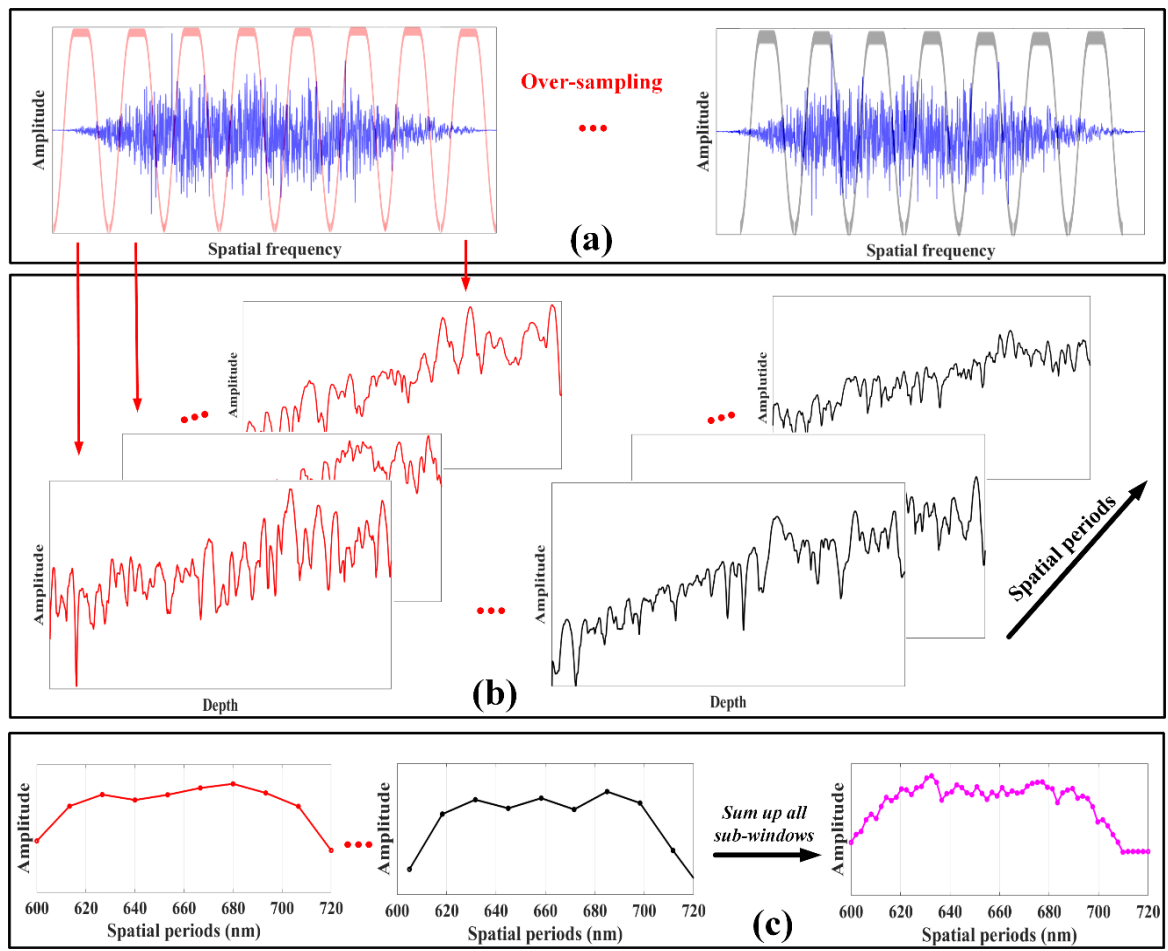

Figure 1. Flowchart for the over-sampling nsOCT signal processing. (a) Decompose axial spatial frequencies into multiple sub-bands and shift the dividing windows for over-sampling the spatial frequencies. (b) FFT of each sub-band and acquire the amplitude along different depths in each particular spatial frequency range. (c) Reconstruct and sum up all the spatial period profiles by taking the energy contribution of each sub-band and then obtain the dominant structural size for each voxel.

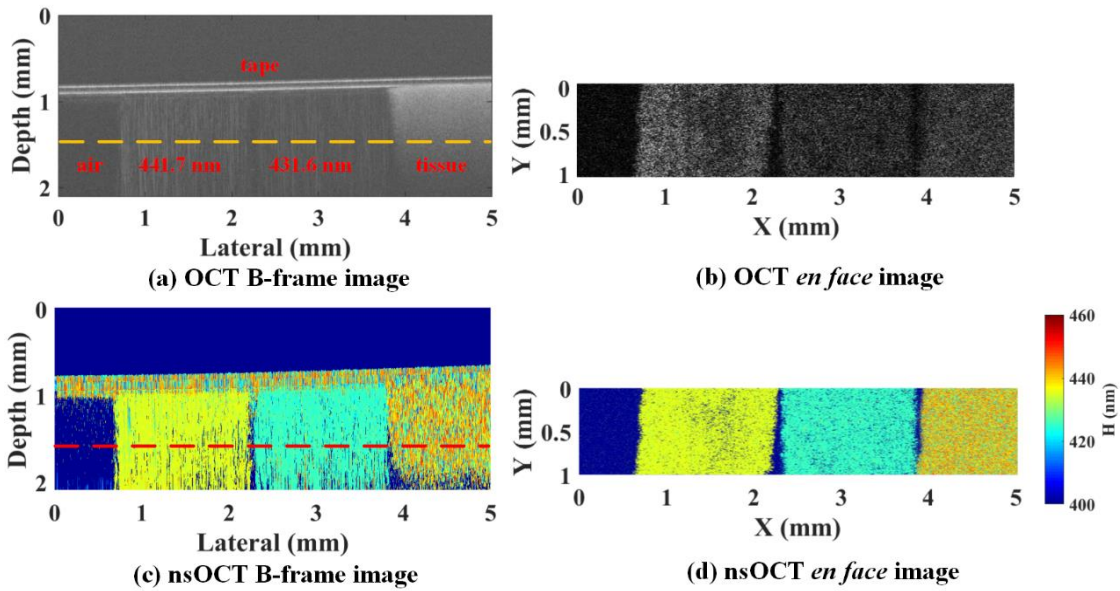

**Figure 2.** Experimental validation of the over-sampling nsOCT method by both B-scan and *en face* images. The two periodicity-different Bragg gratings and a tissue paper were covered under a two-layer tape. (a, b) Conventional intensity-based OCT B-scan and *en face* image at the depth marked by the dashed line. (c, d) The corresponding nsOCT B-scan and *en face* image with spatial periods mapping, visualizing the nanoscale structural differences between the samples. The color bar presents the spatial periods with unit nanometer.

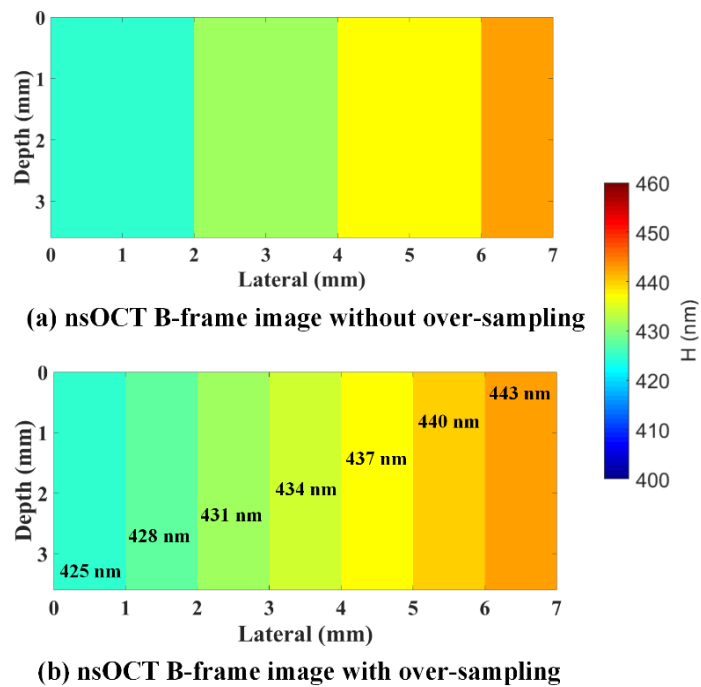

Figure 3. Simulation results of nsOCT without and with over-sampling. (a) nsOCT without over-sampling can detect the difference of 6 nm but cannot distinguish 3 nm difference, (b) nsOCT with over-sampling is able to distinguish the structural difference of 3 nm.

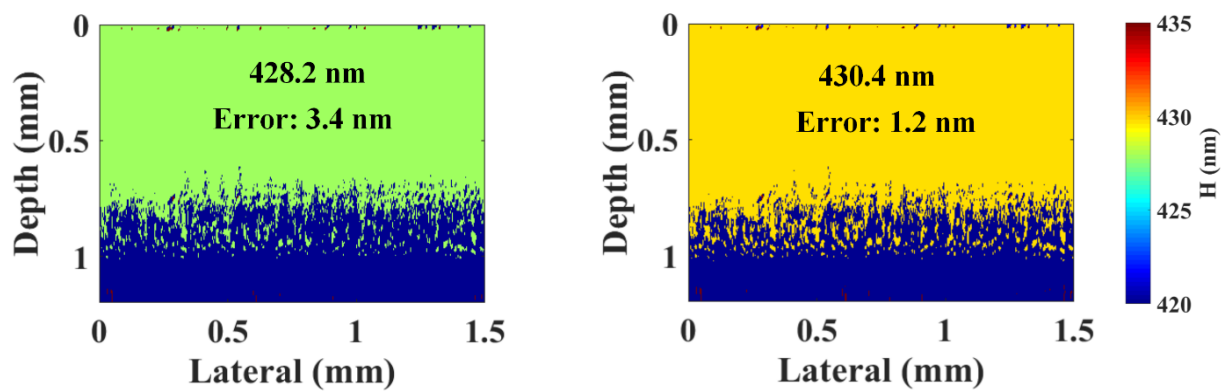

(a) nsOCT image without over-sampling  (b) nsOCT image with over-sampling

Figure 4. Experimental results of the Bragg grating with axial periodicity of 431.6 nm. (a) The nsOCT without over-sampling presents a 428.2 nm structural size. Error is 3.4 nm. (b)The nsOCT with over-sampling shows a 430.4 nm structural size. Error is 1.2 nm. The color bar presents the spatial periods with unit nanometer.

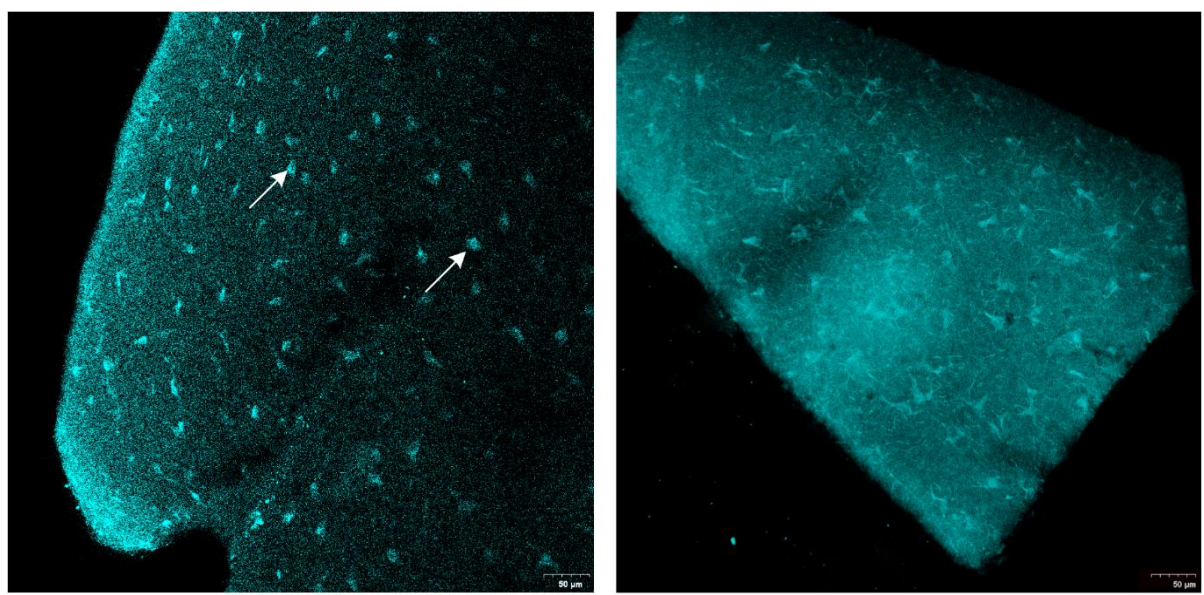

(a) Virgin cornea  (b) Treated cornea

Figure 5. Confocal microscopy images of the virgin and treated cornea. (a) The virgin corneas consists of highly reflective and well-demarcated cell nuclei with an oval shape, indicated by white arrows. (b) The treated cornea is populated with some reflective interconnected stellate structures, containing elongated nuclei and keratocyte apoptotic bodies. The scale bar is 50 μm.

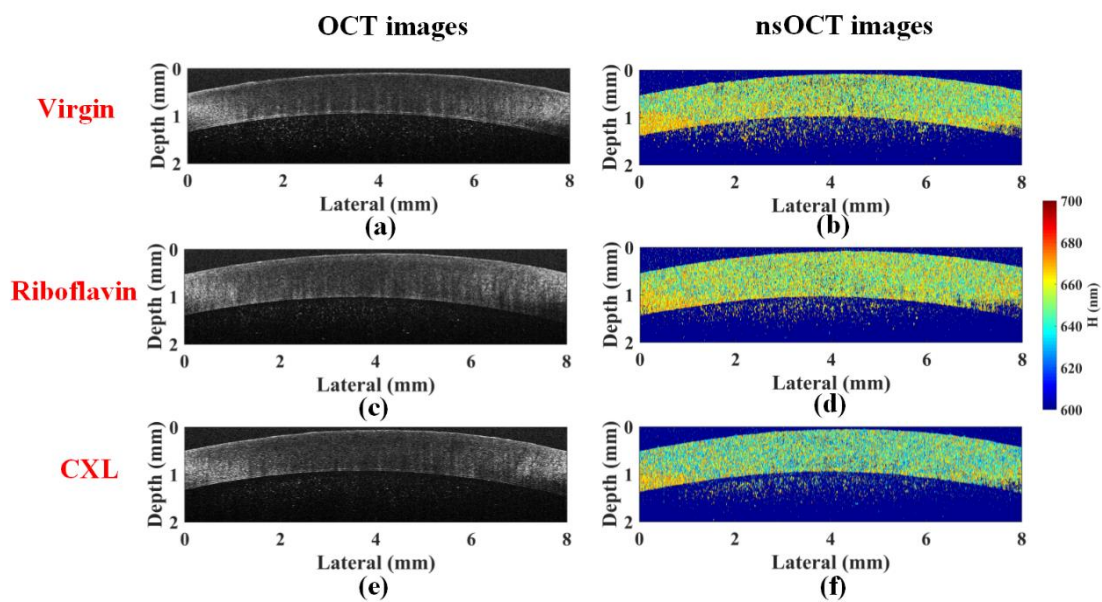

**Figure 6.** B-scans of *ex vivo* bovine cornea during CXL treatment at three treating steps. (a, b) Both conventional OCT and nsOCT images acquired immediately after epithelial removal. (c, d) Images constructed after 30 min riboflavin instillation and (e, f) after 30 min UVA irradiation while finishing the treatment. The color bar presents the spatial periods with unit nanometer.

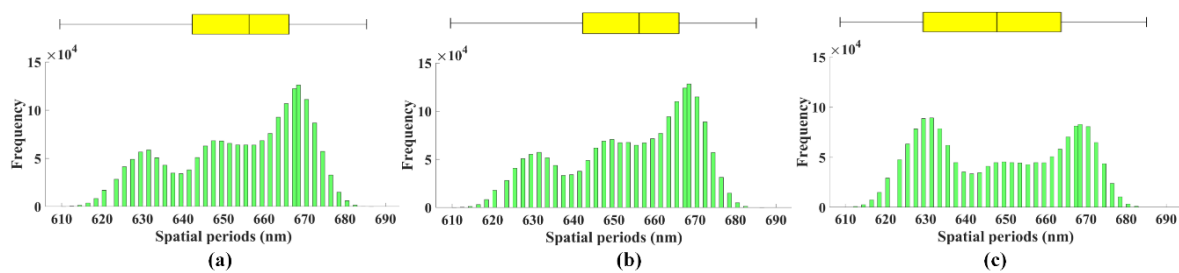

**Figure 7.** Histogram and boxplot of spatial period distributions inside cornea associated with CXL treatment. (a) Virgin cornea with the mean spatial period of 657.1 nm. (b) The cornea after 30 minutes riboflavin instillation with the mean value of 658.1 nm. (c) The crosslinked

cornea with the mean value of 649.6 nm, with the histogram showing a shift towards lower spatial period compared with (a) and (b).

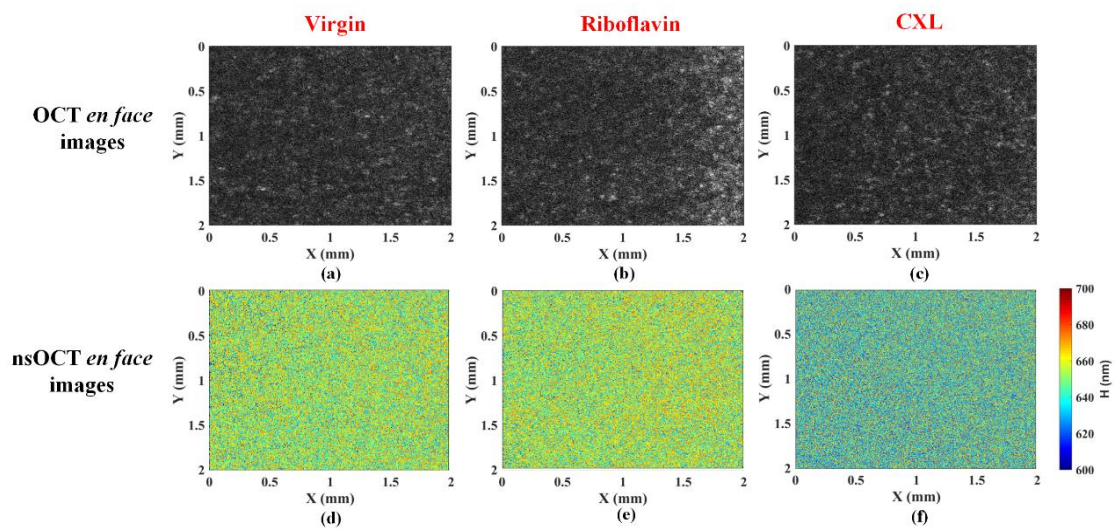

**Figure 8.** Conventional OCT and nsOCT *en face* images at the three key treating steps during CXL treatment. (a, b, c) Conventional intensity-based OCT *en face* images of the cornea after epithelium removal, after 30 min riboflavin instillation and after 30 min UVA illumination, respectively. (d, e, f) The corresponding nsOCT *en face* images with spatial period mapping, presenting the nanoscale structural alterations. The *en face* images were constructed at the depth of ~500 μm. The color bar presents the spatial periods with unit nanometer.

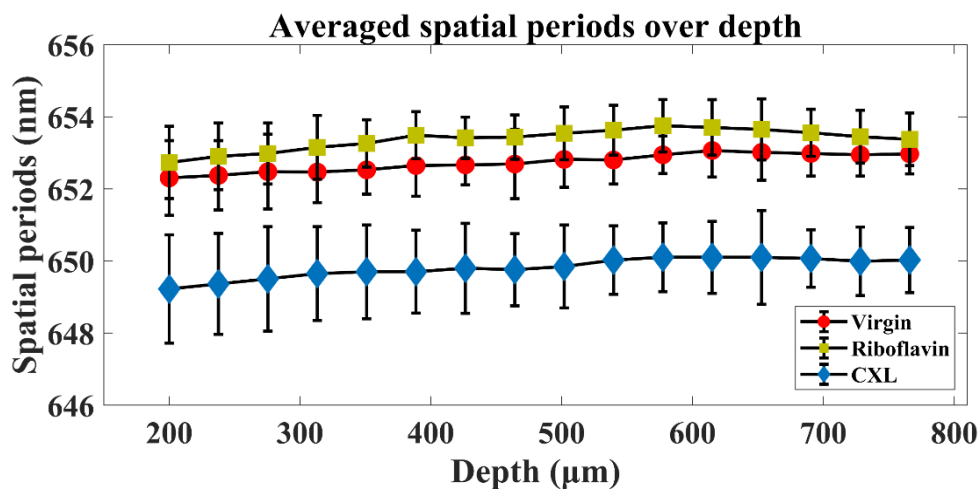

**Figure 9.** Mean spatial periods at various corneal depths for the virgin cornea, the cornea after 30 min riboflavin instillation and the cross-linked cornea, indicated by red, yellow and blue color, respectively.

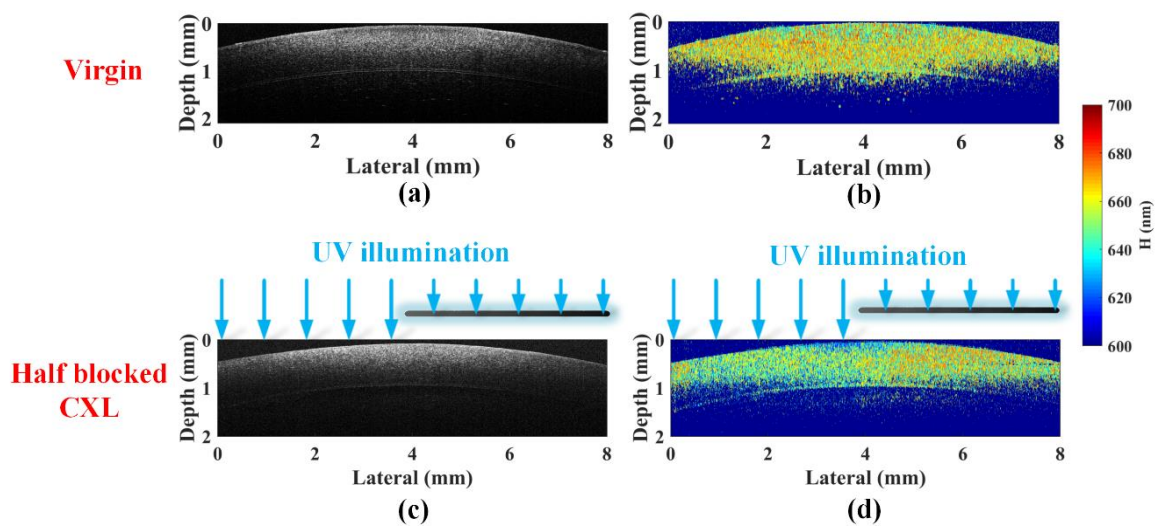

**Figure 10.** A control experiment with the cornea half blocked from UVA irradiation. (a, b) Conventional intensity-based OCT and nsOCT images of the virgin cornea. (c) The conventional OCT image of the half cross-linked cornea using a UVA irradiation block on the right half. (d) The corresponding nsOCT image of the half cross-linked cornea, illustrating the nanoscale structural difference between left half and right half in the same B-scan. The color bar presents the spatial periods with unit nanometer.

**TABLE 1.** The intensity changes between virgin cornea and cross-linked cornea during the CXL treatment

| Bovine corneas | Mean intensity values (dB) | | | |
|---|---|---|---|---|
| | Virgin | Treated | Shift | Percentage |
| S1 | 35.86 | 35.57 | -0.29 | -0.81% |
| S2 | 35.99 | 35.31 | -0.68 | -1.89% |
| S3 | 35.58 | 34.96 | -0.62 | -1.74% |
| S4 | 36.01 | 36.55 | +0.54 | +1.50% |
| S5 | 35.90 | 35.36 | -0.54 | -1.50% |
| S6 | 35.62 | 36.01 | +0.39 | +1.09% |
| S7 | 35.88 | 36.13 | +0.25 | +0.70% |
| S8 | 35.79 | 35.99 | +0.20 | +0.56% |
| S9 | 36.03 | 35.69 | -0.34 | -0.94% |
| S10 | 35.86 | 35.32 | -0.54 | -1.50% |
| Mean | 35.85± 0.15 | 35.69± 0.48 | -0.16± 0.46 | (-0.45± 1.23)% |

**TABLE 2.** The spatial period changes of virgin cornea and cross-linked cornea during the CXL treatment

| Bovine corneas | Mean spatial periods (nm) | | | |
|---|---|---|---|---|
| | Virgin | Treated | Shift | Percentage |
| S1 | 657.1 | 649.6 | -7.5 | -1.14% |
| S2 | 652.1 | 638.5 | -13.6 | -2.09% |
| S3 | 656.3 | 649.2 | -7.1 | -1.08% |
| S4 | 654.3 | 648.1 | -6.2 | -0.95% |
| S5 | 653.2 | 649.6 | -3.6 | -0.55% |
| S6 | 651.7 | 646.9 | -4.8 | -0.74% |
| S7 | 653.6 | 648.9 | -4.7 | -0.72% |
| S8 | 653.9 | 647.3 | -6.6 | -1.01% |
| S9 | 654.3 | 649.6 | -4.7 | -0.72% |
| S10 | 652.6 | 647.3 | -5.3 | -0.81% |
| Mean | 653.9± 1.7 | 647.5± 3.3 | -6.4± 2.8 | (-0.98± 0.43)% |

**Graphical Abstract**

We report the application of over-sampling nano-sensitive optical coherence tomography method for monitoring the nanoscale structural changes inside cornea during corneal cross-linking (CXL) treatment. The proposed method can be implemented using the existing OCT system, without any additional components, to achieve fast, non-invasive and nanoscale sensitivity detection. It can be an indicator in diagnostic assessment associated with CXL treatment, and possibly to be a real-time monitoring tool in clinics.

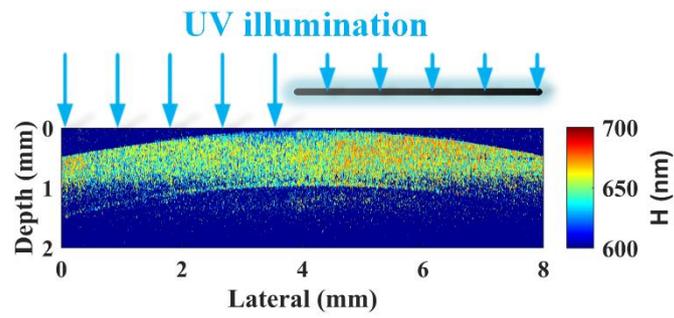